\documentclass{llncs}
\usepackage{t1enc}
\usepackage{rotating}

\begin{document}
\mainmatter              % start of the contributions
\title{{\LARGE ICOOOLPS'2007}\\Second ECOOP Workshop on {\underline
  I}mplementation,  {\underline C}ompilation,   {\underline
  O}ptimization of  {\underline O}bject-{\underline O}riented
  {\underline L}anguages, {\underline P}rograms and {\underline S}ystems} 
\titlerunning{ICOOOLPS'2007}  % abbreviated title (for running head)
\author{Olivier Zendra\inst{1} 
\and Eric Jul \inst{2}
\and Roland Ducournau\inst{3}
\and Etienne Gagnon\inst{4}
\and Richard Jones\inst{5}
\and Chandra Krintz\inst{6}
\and Philippe Mulet\inst{7}
\and Jan Vitek\inst{8}
}

\institute{INRIA-LORIA, France 
\and  DIKU, Denmark 
\and  LIRMM, France
\and  UQAM, Canada
\and  Univ. of Kent, UK
\and  UCSB, USA	
\and  IBM, France
\and  Purdue University, USA	
}

%
%%%\authorrunning{Olivier Zendra}   % abbreviated author list (for running head)
%
%%%% modified list of authors for the TOC (add the affiliations)
%%%\tocauthor{Oliver Zendra (INRIA - LORIA)}
%
%%%\institute{INRIA-Lorraine / LORIA, Building C, \\
%%%615 Rue du Jardin Botanique, BP 101, \\
%%%54602 Villers-L\`{e}s-Nancy Cedex,\\
%%% FRANCE\\
%%%\email{Olivier.Zendra@loria.fr}\\ 
%%%\texttt{http://www.loria.fr/\homedir zendra}
%%%}

\maketitle              % typeset the title of the contribution

\begin{abstract}
ICOOOLPS'2007 was the second edition of the ECOOP-ICOOOLPS workshop.
ICOOOLPS intends to bring researchers and practitioners
both from academia and industry together, with a spirit of openness,
to try and identify and begin to address the numerous and very varied
issues of optimization. 
After a first successful edition, this second one put a stronger emphasis on 
exchanges and discussions amongst the participants, progressing on the bases
set last year in Nantes.

The workshop attendance was a success, since the 30-people limit we had set
was reached about 2 weeks before the workshop itself.
Some of the discussions (e.g .annotations) were so successful that they would
required even more time than we were able to dedicate to them. 
That's one area we plan to further improve for the next edition.
\end{abstract}
%

%%%%%%%%%%%%%%%%%%%%%%%%%%%%%%%%%%%%%%%%%%%%%%%%%%%%%%%%%%%%%%%%%%%%%%
\section{Objectives and call for papers}

Programming languages, especially object-oriented ones, are pervasive and 
play a significant role in computer science and engineering life. 
They sometime appear as
ubiquitous and completely mature. However, despite a large number of
works, there is still a clear need for solutions for efficient
implementation and compilation of OO languages in various application
domains ranging from embedded and real-time systems to desktop 
systems. 

The ICOOOLPS workshop series thus aims to address this crucial issue of
optimization in OO languages, programs and systems. It intends to do
so by bringing together researchers and practitioners working in the
field of object-oriented languages implementation and
optimization. Its main goals are identifying fundamental bases and key
current issues pertaining to the efficient implementation, compilation
and optimization of OO languages, and outlining future challenges and
research directions. 

\smallskip

Topics of interest for ICOOOLPS include but are not limited to:
\begin{itemize}
    \item  implementation of fundamental OOL features:
       \begin{itemize}
          \item inheritance (object layout, late binding, subtype test...)
          \item genericity (parametric types)
          \item memory management
       \end{itemize}
    \item  runtime systems:
       \begin{itemize}
          \item compilers
          \item linkers
          \item virtual machines
       \end{itemize}
    \item  optimizations:
       \begin{itemize}
          \item static and dynamic analyses
          \item adaptive virtual machines
       \end{itemize}
    \item  resource constraints:
       \begin{itemize}
          \item real-time systems
          \item embedded systems (space, low power)...
       \end{itemize}
    \item  relevant choices and tradeoffs:
       \begin{itemize}
          \item constant time vs. non-constant time mechanisms
          \item separate compilation vs. global compilation
          \item dynamic loading vs. global linking
          \item dynamic checking vs. proof-carrying code
          \item annotations vs. no annotations
       \end{itemize}
\end{itemize}

This workshop thus tries to identify
fundamental bases and key current issues pertaining to the efficient
implementation and compilation of languages, especially OO ones, in order
to spread them further amongst the various computing systems. It is also
intended to extend this synthesis to encompass future challenges and
research directions in the field of OO languages implementation and
optimization. 

Finally, as stated from the very beginning and the very first edition in 
Nantes in 2006, ICOOOLPS is intended to be a recurrent workshop in ECOOP.
Since the feedback from first year attendants was very positive, 
this second edition was set up. We organizers integrated most of the 
suggestions for improvements made in 2006, so as to further improve the 
workshop. The main adaptation was that less time was given to presentations, 
in order to free extra time for discussions.

In order to increase bases on which the discussions could be based
and to keep them focused, each prospective participant was 
encouraged to submit either a short paper describing ongoing work or a
position paper describing an open issue, likely solutions, drawbacks
of current solutions or alternative solutions to well known problems. 
Papers had to be written in English and their final version could not
exceed 8 pages in LNCS style (4 pages recommended).

%%%%%%%%%%%%%%%%%%%%%%%%%%%%%%%%%%%%%%%%%%%%%%%%%%%%%%%%%%%%%%%%%%%%%%
\section{Organizers}

\medskip

\begin{tabbing}
\hspace*{5mmm}\=\hspace*{2cm}\=\hspace{2cm}\=\kill
{\bf Olivier ZENDRA (chair)}, \>\>\>INRIA-LORIA, Nancy, France.\\
 \>Email:   \> {\tt olivier.zendra@inria.fr}\\
 \>Web:     \> {\tt http://www.loria.fr/\homedir zendra}\\
 \>Address: \> INRIA / LORIA\\
 \>         \> 615 Rue du Jardin Botanique\\
 \>	    \> BP 101\\
 \>	    \> 54602 Villers-Lès-Nancy Cedex, FRANCE\\
\end{tabbing}
\vspace{-5mm}
	Olivier Zendra is a full-time permanent computer science
	researcher at INRIA~/~LORIA, in Nancy, France. His research
	topics cover compilation, optimization and automatic memory
	management. He worked on the compilation and optimization of
	object-oriented languages and was one of the two people who
	created and implemented SmartEiffel, The GNU Eiffel Compiler
	(at the time SmallEiffel). His current research topics and 
	application domains are program analysis, compilation, memory 
	management and embedded
	systems, with a specific focus on low energy.

\begin{tabbing}
\hspace*{5mmm}\=\hspace*{2cm}\=\hspace{2cm}\=\kill
{\bf Eric JUL (co-chair)}, \>\>\>DIKU, Copenhagen, Denmark.\\
 \>Email:   \> {\tt eric@diku.dk}\\
 \>Web:     \> {\tt http://www.diku.dk/\homedir eric}\\
 \>Address: \> DIKU\\
 \>         \> Universitetsparken 1\\
 \>	    \> DK-2100 København Ø, DANMARK\\
\end{tabbing}
\vspace{-5mm}
Eric Jul is Professor of Computer Science at the University of
Copenhagen and head of the Distributed Systems Group.
He is one of the principal designers of the distributed, object-oriented
language Emerald. He implemented fine-grained object mobility in
Emerald. His current research is in Grid Computing. He is currently
Vice-President of AITO.

\medskip

\begin{tabbing}
\hspace*{5mmm}\=\hspace*{2cm}\=\hspace{3cm}\=\kill
{\bf Roland DUCOURNAU}, \>\>\>~~~~~~~ LIRMM, Montpellier, France.\\
 \>Email:   \> {\tt ducour@lirmm.fr}\\
 \>Web:     \> {\tt http://www.lirmm.fr/\homedir ducour}\\
 \>Address: \> LIRMM,\\
 \>         \> 161, rue Ada\\
 \>	    \> 34392 Montpellier Cedex 5, FRANCE \\
\end{tabbing}
\vspace{-5mm}
	Roland Ducournau is Professor of Computer Science at the
	University of Montpellier. In the late 80s, while with Sema
	Group, he designed and developed the YAFOOL language, based on
	frames and prototypes and dedicated to knowledge based
	systems. His research topics focuses on class specialization
	and inheritance, especially multiple inheritance. His recent
	works are dedicated to implementation of OO languages. 

\medskip

\begin{tabbing}
\hspace*{5mmm}\=\hspace*{2cm}\=\hspace{3cm}\=\kill
{\bf Etienne GAGNON}, \>\>\>UQAM, Montréal, Québec, Canada.\\
 \>Email:   \> {\tt egagnon@sablevm.org}\\
 \>Web:     \> {\tt http://www.info2.uqam.ca/\homedir egagnon}\\
 \>Address: \> Département d'informatique\\
 \>         \> UQAM \\
 \>	    \> Case postale 8888, succursale Centre-ville \\
 \>	    \> Montréal (Québec) Canada / H3C 3P8\\
\end{tabbing}
\vspace{-5mm}
	Etienne Gagnon is a Professor of Computer Science at
	Université du Québec à Montréal (UQAM) since 2001. Etienne has
	developed the SableVM portable research virtual machine for
	Java, and the SableCC compiler framework generator. His
	research topics include language design, memory management,
	synchronization, verification, portability, and efficient
	interpretation techniques in virtual machines. 

\medskip

\begin{tabbing}
\hspace*{5mmm}\=\hspace*{2cm}\=\hspace{3cm}\=\kill
{\bf Richard JONES}, \>\>\>University of Kent, Canterbury, UK.\\
 \>Email:   \> {\tt R.E.Jones@kent.ac.uk}\\
 \>Web:     \> {\tt http://www.cs.kent.ac.uk/\homedir rej}\\
 \>Address: \> Richard Jones, Reader in Computer Systems,\\
 \>         \> Computing Laboratory, \\
 \>	    \>  University of Kent at Canterbury, \\
 \>	    \> Canterbury CT2 7NF, UK\\
\end{tabbing}
\vspace{-5mm}
Richard Jones is Reader in Computer Systems and Deputy Director of the
Computing Laboratory at the University of Kent, Canterbury.  He leads
the Systems Research Group.
He is best known for his work on garbage collection: his monograph
Garbage Collection remains the definitive book on the subject.  His memory
management research interests include techniques for avoiding space
leaks, scalable yet complete garbage collection for distributed systems,
flexible techniques for capturing traces of program behaviour, and heap
visualisation.
He was made a Distinguished Scientist of the Association for Computer
Machinery (ACM) in 2006 and awarded an Honorary Fellowship at the
University of Glasgow in 2005.

\medskip

\begin{tabbing}
\hspace*{5mmm}\=\hspace*{2cm}\=\hspace{3cm}\=\kill
{\bf Chandra KRINTZ}, \>\>\>UC Santa Barbara, CA, USA.\\
 \>Email:   \> {\tt ckrintz@cs.ucsb.edu}\\
 \>Web:     \> {\tt http://www.cs.ucsb.edu/\homedir ckrintz}\\
 \>Address: \> University of California\\
 \>         \> Engineering I, Rm. 1121\\
 \>	    \> Department of Computer Science\\
 \>	    \> Santa Barbara, CA 93106-5110, USA\\
\end{tabbing}
\vspace{-5mm}

	Chandra Krintz is an Assistant Professor at the University of
	California, Santa Barbara (UCSB); she joined the UCSB faculty
	in 2001. Chandra's research interests include automatic and
	adaptive compiler and virtual runtime techniques for
	object-oriented languages that improve performance and
	increase battery life. In particular, her work focuses on
	exploiting repeating patterns in the time-varying behavior of
	underlying resources, applications, and workloads to guide
	dynamic optimization and specialization of program and system
	components. 

\medskip

\begin{tabbing}
\hspace*{5mmm}\=\hspace*{2cm}\=\hspace{3cm}\=\kill
{\bf Philippe MULET}, \>\>\>IBM, Saint-Nazaire, France.\\
 \>Email:   \> {\tt philippe\_mulet@fr.ibm.com}\\
 \>Address: \> IBM France - Paris Laboratory\\
 \>         \> 69, rue de la Vecquerie\\
 \>	    \> 44600 Saint-Nazaire, France\\
\end{tabbing}
\vspace{-5mm}

	Philippe Mulet is the lead for the Java Development Tooling
	(JDT) Eclipse subproject, working at IBM since 1996; he is
	currently located in Saint-Nazaire (France). In late 1990s,
	Philippe was responsible for the compiler and codeassist tools
	in IBM Java Integrated Development Environments (IDEs):
	VisualAge for Java standard and micro editions. Philippe then
	became in charge of the Java infrastructure for the Eclipse
	platform, and more recently of the entire Java tooling for
	Eclipse. Philippe is a member of the Eclipse Project
	PMC. Philippe is also a member of the expert group on compiler
	API (JSR199), representing IBM. His main interests are in
	compilation, performance, scalability and meta-level
	architectures.

\medskip

\begin{tabbing}
\hspace*{5mmm}\=\hspace*{2cm}\=\hspace{3cm}\=\kill
{\bf Jan VITEK}, \>\>\>Purdue Univ., West Lafayette, IN, USA.\\
 \>Email:   \> {\tt jv@cs.purdue.edu}\\
 \>Web:     \> {\tt http://www.cs.purdue.edu/homes/jv}\\
 \>Address: \> Dept. of Computer Sciences\\
 \>         \> Purdue University\\
 \>	    \> West Lafayette, IN 47907, USA\\
\end{tabbing}
\vspace{-5mm}

	Jan Vitek is an Associate Professor in Computer Science at
	Purdue University. He leads the Secure Software Systems
	lab. He obtained his PhD from the University of Geneva in
	1999, and a MSc from the University of Victoria in
	1995. Prof. Vitek research interests include programming
	language, virtual machines, mobile code, software engineering
	and information security.

%%%%%%%%%%%%%%%%%%%%%%%%%%%%%%%%%%%%%%%%%%%%%%%%%%%%%%%%%%%%%%%%%%%%%%
\section{Participants}

ICOOOLPS attendance was limited to 30 people for technical reasons.
Unlike in the 2006 edition, it was not mandatory for ICOOOLPS 2007 to submit 
a paper to participate. We indeed intended to further open the discussion by
making the attendance easier, and had learned from the numerous walk-ins 
during ICOOOLPS 2006.
The 30-people limit was reached about 2 weeks before the workshop itself, 
which lead us to put a note on the website to stop new registrations.

Finally, 27 people from 12 countries --- up from 22 people from 8 countries 
in 2006 --- attended this second edition, which is an encouraging 
sign of an increasing audience for ICOOOLPS.
These attendants are  listed in table \ref{tab:attendees}.

\bigskip 

\hspace{-17mm}
\begin{table}
\begin{sideways}
\begin{tabular}{|l|l|l|l|l|}
\hline
First name & NAME & Affiliation & Country & Email \\ 
\hline
\hline
Philippe& ALTHERR&	Google&	Switzerland& {\tt paltherr@google.com}\\
Maurizio& CIMADAMORE&	DEIS, Università di Bologna&	Italy& {\tt Maurizio.Cimadamore@unibo.it}\\
Marcus& DENKER&	Ubiversity of Bern&	Switzerland& 
{\tt denker@iam.unibe.ch}\\
Iulian& DRAGOS&	EPFL&	Switzerland& {\tt iulian.dragos@epfl.ch}\\
Gilles& DUBOCHET&	EPFL - LAMP&	Switzerland& 
{\tt Gilles.Dubochet@epfl.ch}\\
Burak& EMIR&	EPFL&	Switzerland& {\tt Burak.Emir@gmail.com}\\
Michael& FRANZ&	UC Irvine&	USA& {\tt franz@uci.edu}\\
Etienne & GAGNON &  UQAM & Canada & {\tt egagnon@sablevm.org} \\
Michael& HAUPT&	Hasso-Plattner-Institut, Univ. Potsdam& 	Germany& {\tt michael.haupt@hpi.uni-potsdam.de}\\
Raymond& HU&	Imperial College, London&	United Kingdom& {\tt rh105@doc.ic.ac.uk}\\
Christine& HUNDT&	TU-Berlin&	Germany& {\tt resix@cs.tu-berlin.de}\\
Maha& IDRISSI AOUAD&	INRIA / LORIA&	France& {\tt Maha.IdrissiAouad@loria.fr}\\
Eric & JUL & DIKU & Denmark & {\tt eric@diku.dk} \\
Stéphane& MICHELOUD&	EPFL&	Switzerland&{\tt Stephane.MICHELOUD@epfl.ch}\\
Anders Bach& NIELSEN&	University of Århus&	Denmark& {\tt abachn@daimi.au.dk}\\
Meir& OVADIA&	Cadence&	Israel& {\tt meiro@cadence.com}\\
Laurent& PLAGNE&	EDF R\&D& France& {\tt Laurent.Plagne@edf.fr}\\
Andreas& PRIESNITZ&	Chalmers University of Technology&	Sweden&{\tt priesnit@cs.chalmers.se}\\
Yannis& SMARAGDAGKIS&	University of Oregon&	USA& {\tt yannis@cs.uoregon.edu}\\
Alexander& SPOON&	EPFL&	Switzerland& {\tt lex@lexspoon.org}\\
Witawas& SRISA-AN&	University of Nebraska - Lincoln&	USA& {\tt witty@cse.unl.edu}\\
Darko& STEFANOVIC&	University of New Mexico&	USA& {\tt darko@cs.unm.edu}\\
Jan& SZUMIEC&	Cracow University of Technology&	Poland& {\tt jps@wieik.pk.edu.pl}\\
Howard& THOMSON&	UKUUG Council&	United Kingdom& {\tt howard.thomson@dial.pipex.com}\\
Stijn& TIMBERMONT&	Vrije Universiteit Brussel&	Belgium& {\tt stimberm@vub.ac.be}\\
Jan & VITEK & Purdue Univ. & USA & {\tt v@cs.purdue.edu} \\
Olivier & ZENDRA & INRIA-LORIA &  France & {\tt Olivier.Zendra@loria.fr} \\ 
\hline
\end{tabular}
\end{sideways}
\caption{ICOOOLPS 2007 list of attendees}
\label{tab:attendees}
\end{table}

%%%%%%%%%%%%%%%%%%%%%%%%%%%%%%%%%%%%%%%%%%%%%%%%%%%%%%%%%%%%%%%%%%%%%%
\section{Contributions}

The presentations and discussions at ICOOOLPS 2007 were organized in 
4 sessions: annotations vs. no annotation, lookup and dispatch mechanisms, 
miscellaneous implementation issues and continuations and synchronizations.

Here are the main contributions for the sessions. More details (papers, presentations slides, etc.) are available from {\tt http://icooolps.loria.fr}.
They are reported here in a lively an rather informal way, so as to keep some of the spontaneity of the workshop, with of course extra organization.

%%%%%%%%%%%%%%%%%%%%%%%%%%%%%%%%%%%%%%%%%%
\subsection{Annotations vs. no annotation}
\label{sec:annotations}

This first technical session was a discussion-only one, chaired by Olivier 
Zendra, who introduced it by a talk synthesizing the contributions of 
ICOOOLPS 2006 discussion "written down in code vs. inferred".
It was a very lively and interesting discussion, with a lot of attendees participating. Unfortunately, to respect the schedule, we had to stop the discussion before it was over.
This first indicates this discussion topic is still open and should probably be continued in 2008, then that discussion times should be even longer and/or more flexible.

\bigskip

A quote from last year stated that "Annotations are too serious to be 
left to developers". But this triggers the question "And what about code ?!"

Some answers pointed that there is room for the compiler to do consistency checking.
Others argued it was better to let people do their own mistakes, since that's part of the learning process. It was objected that this reasoning, pushed to the extreme, could lead to directly writing assembly code. Everyone agreed that of course we still need higher level because we want people be more productive.

The issue was raised whether we actually needed different levels of annotations.
One level would we the "How-level", where we express how things are done. This is very useful for optimization. Not so many people in the room considered this level appealing to them, though.
Another level would be the "What-level", where we express properties (eg security) of the program, algorithms, ie. what has to be done to some extent. Many people in the room considered this level appealing to them.

But a flag was waived: annotations that change the meaning of a program are just ... code !
So annotations should not change the semantic of a program, otherwise we obtain a new language. Annotations, to remain genuine ones, should be intrinsically optional: they should be {\em hints}.
Annotations can be constraints. They thus express domain-specific things and pertain to checking. However, annotations should not grow so much as to have their own type system, otherwise this makes the program much more complex.

A very interesting point was that we may need different hints, for different uses, for different people (annotations for security, for speed optimization, for .... ?)
So one remarked that maybe they should stay {\em outside} the code of the program itself.
We could have source (code) files and annotations files, each pertaining to a specific domain.

But wouldn't it be better to be able to modify the language easily (extension, reflexivity...) ? That could be an opening question for next year !

Reflexive annotations (with run-time changes) were mentioned, but the discussion did not go very far on this.

%%%%%%%%%%%%%%%%%%%%%%%%%%
\subsection{Lookup, dispatch mechanisms}
\label{sec:lookup}

The second session, chaired by Eric Jul, consisted of 2 paper presentations, 
one insightful introductory talk by Eric on AbCons, and a discussion.
This session topic was a brand new one from this year. 

\bigskip

The first paper, "One method at a time is quite a waste of time", by Andreas Gal, Michael Bebenita and Michael Franz (University of California, Irvine, USA), made a very convincing case that optimizing on a per method basis is not a good granularity level. Instead their compiler optimizes on at the granularity of hot traces, especially for loops.

The second paper, "Type feedback for bytecode interpreters", by Michael Haupt, Robert Hirschfeld (Univ. of Potsdam, Germany) and Marcus Denker (Univ. of Bern, Switzerland), explained the advantages pertaining to the use of polymorphic inline caches (PICs) in interpreters, and some implementation details in Squeak Smalltalk.

After these nice research works and the introduction on AbCons by Eric,
 the discussion itself unfortunately did
not really catch up, it seems. Things were probably not mature enough. 
It is also possible that the attendees were not concerned by this kind of 
implementation "details"... Maybe we could check this for next year (survey ?).
The timing --- just before lunch --- may also have had an impact.

A few points of interest nonetheless emerged:
\begin{itemize}
 \item Lookup can be implemented in many different ways.
 \item Lookup tends to increase memory size. This is not too good for caches, hence performance.
 \item Similarly, lookup tends to increase register pressure, with again a negative impact on performance.
 \item There was some discussion about the use of fat pointers, to reduce the cost of lookup. Some participants argued that fat pointers are too expensive.
 \item Most calls can be solved statically, hence alleviating the need for (run-time) lookup. Of course, this may imply whole system analysis, possibly at link time.
\end{itemize}

%%%%%%%%%%%%%%%%%%%%%%%%%%
\subsection{Miscellaneous implementation issues}
\label{sec:implementation}

This third session, chaired by Eric Jul, begun the afternoon with three papers.

\bigskip

Titled "A Survey of Scratch-Pad Memory Management Techniques for low-power and -energy", the first paper by Maha Idrissi Aouad (Univ. Henri-Poincaré, Nancy, France) and Olivier Zendra (INRIA-LORIA, Nancy, France) presented various existing SPM (scratch-pad memory) management techniques aimed at low-power. It mostly focused on optimal placement of data according to existing techniques and outlined unexplored directions.

The second paper, "Language and Runtime Implementation of Sessions for Java" by 
Raymond Hu, Nobuko Yoshida (Imperial College, London, United Kingdom) and Kohei Honda (Univ. of London, United Kingdom), explained how session types could provide type-safe communications in Java. An implementation validating this was shown, with important protocol and communications points detailed.

Finally, "Ensuring that User Defined Code does not See Uninitialized Fields" by 
Anders Bach Nielsen (Univ. of Aarhus, Denmark) was the third and last paper of this sessions. It discussed some of the problems and solutions found in implementing gbeta, a generalization of the BETA language. This ongoing work focused on a smart handling of object initialization so as to guarantee that user code only uses fully initialized object, thus strengthening the type system promises.

%%%%%%%%%%%%%%%%%%%%%%%%%%
\subsection{Continuations and synchronizations}
\label{sec:synchronization}

This fourth session of ICOOOLPS 2007 was chaired by Etienne Gagnon and comprised one paper, one detailed presentation by Etienne on fat locks and Java synchronization and a discussion.
It continued ICOOOLPS 2006 unfinished discussion about threads in Java.

\bigskip

The paper in this session was presented by Iulian Dragos (EPFL, Switzerland), Antonio Cunei and Jan Vitek (Purdue Univ., USA). Titled "Continuations in the Java Virtual Machine", it was an introduction to the nontrivial addition of first-class continuation in a Java VM. It outlined the issues such an addition raises, studying interactions with existing features of the Java language such as exceptions, threads, security model and garbage collector.

After a very detailed and complete talk on "Keeping fat locks on a diet, eager deadlock detection, and looking beyond the current Java synchronization model" by Etienne, the discussion on "Java threads and synchronization model." took place.

This was a follow-up and extension to last year's discussion "Do (Java) threads make sense ?". This topic sparked a lot of interest, unlike last year, which indicates that the topic had somehow matured in participants minds.

The current statu quo is "rely on the developer" to express and manage concurrency/synchronization.
However, Java was about protecting programmers from themselves. Is it really still the case with threads and synchronization as done in Java ?
Threads are not part of the language in Java, but the "synchronized" keyword is. Shouldn't they both be part of the language ? The current situation is somewhat unbalanced.

We then considered what was in the future. Cooperative synchronization ?
Synchronization is harder than GC (Garbage collection): indeed automating synchronization is not possible, it is part of the semantics (which is not the case for a GC's work).
Synchronization is akin to parallel programming. It's an unsolved problem.
On a high level, writing a language that prevents deadlocks (or tells you there are none) would be great. But isn't it like solving the halting problem ? That's not a promising path...

Once again, participants asked whether Java threads were really useful.
Indeed, threads and their synchronization seem very low level. But to go lower level than Java, we have C... Shared memory and parallelism is ugly but convenient for scientific programming.

The actual problem for developers is to express that they want to use parallelism, not how.
On a higher level, we have parallel programming, join, merge...

Would "actors" and asynchronous message sending be appropriate ?

Overall, the consensus seems to be that threads and synchronization in Java is flawed, not at the appropriate level. Higher-level means should be provided to express these concerns. Those who need lower-level or very fine control of things should rely on going through C code.

%%%%%%%%%%%%%%%%%%%%%%%%%%%%%%%%%%%%%%%%%%%%%%%%%%%%%%%%%%%%%%%%%%%%%%
\section{Conclusion}
\label{sec:conclusions}

This second edition of ICOOOLPS was a successful successor to ICOOOLPS 2006, where it had been decided ICOOOLPS should go on recurrently, on a yearly basis.
This year, we managed to increase the audience of ICOOOLPS, gathering
27 people from 12 countries --- up from 22 people from 8 countries 
in 2006 --- from academia and industry, researcher as well as practitioners.
This clearly bides well for the future and the building of a small, informal,
community.

\smallskip

A number of positive aspects can be mentioned about ICOOOLPS 2007.

First, this year, the workshop was officially open to anyone, not only
authors/speakers. This was coherent with the fact that an ECOOP workshop aims
at fostering discussions and exchanges, and the fact we had had many 
unregistered (but welcome) walk-ins in 2006.

Thanks to our correct forecast for a larger attendance, this year 
the room allocated by the ECOOP organizers was able to comfortably host 
all the attendants.

The name tags for attendants were also a small but welcome improvement.

\smallskip 
On a more scientific level, once again thanks to the skills of the speakers and active participation of the attendants, the discussions were lively, open-minded and allowed good
exchanges. 
We had allocated more time for discussions than last year, but it was barely enough.

Another encouraging aspect is that some discussions (annotations, Java threads) recurred from 2006, which shows there is interesting work to be done in these areas. Furthermore, the fact that the discussion on Java threads, which did not caught up in 2006, was successful this year, indicates that some topics are maturing.

As we had mentioned last year identifying the main challenges for optimization is not  that easy, if only because optimizations for
object-oriented languages come in variety of contexts with very
different constraints (embedded, real-time, dynamic, legacy...) hence different optimizations criteria (speed, size, memory footprint, energy...).
One thing that emerged more clearly in this second edition is the fact that some of our concerns extend beyond object-oriented languages (to functional languages, for example).
Another important point is that to optimize, it is difficult to consider separately implementation and language design, or at least specifications.
In this respect, the consensus we reached in the workshop that threads and synchronization in Java are flawed and not at the appropriate level is an interesting outcome.

%%%%%%%%%%%%%%%%%%%%%%%%%%%%%%%%%%%%%%%%%%%%%%%%%%%%%%%%%%%%%%%%%%%%%%
\section{Perspectives: ICOOOLPS future}
\label{sec:perspectives}

The perspectives for the ECOOP-ICOOOLPS workshop are very good. 
When surveyed at the very end of the workshop, 16 attendees amongst the 
18 still present intended to come next year.
We are thus very confident for ICOOOLSP 2008 to happen, in Cyprus.

Like every year, we try to draw lessons from each edition to further improve the following ICOOOLPS editions. This year, we noted several aspects to improve, amongst which the main ones are:

\begin{itemize}
    \item  This year, we had shorter presentations and longer discussions than in 2006. That was good. But in 2008 we should {\em devote  even more time to discussions}, with even shorter presentations: the purpose of a workshop is not papers, but brainstorming. Presentations should be 10 minutes {\em max} + 10 minutes for questions.
    \item We must be {\em very strict with presentations times}, and not hesitate to stop a speaker who's exceeding her/his time.
    \item The {\em papers} do have to be available on the website {\em before} the workshop.
    \item Session report drafts should be written during a session
    (papers and talks) and maybe briefly discussed at the end of each
    session (not after the workshop).
    \item Prior registration with the workshop organizers, like in ICOOOLPS 2006, is better. It helps keeping track of attendants, gathering their topics of interest, etc.
    \item We have to provide {\em a list of suggested discussion topics} at registration time, so that attendees can vote for them (or suggest new ones). Having discussion time open for topics suggested during the workshop did not work very well in 2007.
\end{itemize}

Of course, some of these points put an increased burden on the organizers, but are key to an even more successful and enjoyable workshop.

We also intend to selectively enlarge the audience to other --- possibly non-OO --- communities who face the same kind of issues as the one we focus on in ICOOOLPS.

%%%%%%%%%%%%%%%%%%%%%%%%%%%%%%%%%%%%%%%%%%%%%%%%%%%%%%%%%%%%%%%%%%%%%%
\section{Background}

In order to provide a fixed access point for ICOOOLPS related matters, the
web site for the workshop is maintained at {\tt
http://icooolps.loria.fr}. 
All the papers and presentations done for ICOOOLPS'2007 are freely
available there.


\nocite{*} 

\bibliographystyle{plain}

\bibliography{icooolps_2007_WS_report}

\end{document}